\documentclass[preprint,review,12pt]{elsarticle}
\usepackage{graphicx}
\usepackage{amssymb}
\journal{Solid State Communications}

\begin{document}
\begin{frontmatter}
\title{Superconducting phase transition in~YNiGe$_3$, a non-$f$-electron reference to~the unconventional superconductor CeNiGe$_3$}
\author{Adam P. Pikul\corref{pikul}}
\author{Daniel Gnida}
\address{Institute of Low Temperature and Structure Research, Polish Academy of Sciences, P Nr 1410, 50-950 Wroc{\l}aw 2, Poland}
\cortext[pikul]{Corresponding author. Phone: +48 71 3435021, Fax: +48 71 3441029, E-mail:
A.Pikul@int.pan.wroc.pl}

\begin{abstract}
A polycrystalline sample of YNiGe$_3$, being a non-magnetic isostructural counterpart to the
unconventional pressure-induced superconductor CeNiGe$_3$, was studied by means of specific heat
and electrical resistivity measurements at temperatures down to 360~mK and in magnetic fields up to
500~Oe. The compound was found to exhibit an ambient-pressure superconductivity below $T_{\rm
c}$~=~0.46~K. The superconducting state in YNiGe$_3$ is destroyed by magnetic field of the order of
500~Oe.
\end{abstract}


\end{frontmatter}

\newpage

\section{Introduction}

The compound CeNiGe$_3$, crystallizing in the orthorhombic SmNiGe$_3$ type
structure,\cite{salamakha} is an antiferromagnetically ordered Kondo lattice with the N{\'e}el
temperature $T_{\rm N}$~=~5.5~K and the characteristic Kondo temperature $T_{\rm K} \sim T_{\rm
N}$~\cite{pikul}. Upon applying hydrostatic pressure, $T_{\rm N}$ increases up to about 8~K at
$P_{\rm max} \approx$~3~GPa, rapidly decreases at higher pressure, and finally is suppressed to
zero at a~critical pressure $P_{\rm c} \approx$~5.5~GPa. Most importantly, around this quantum
critical point the compound becomes superconducting below about 0.48~K. Analysis of the temperature
variation of its electrical resistivity in the normal state revealed an unconventional nature of
the superconductivity in CeNiGe$_3$.~\cite{nakashima1,nakashima2} More detailed investigations of
the compound showed that the superconductivity emerges in CeNiGe$_3$ not only in the critical
region, but also deeply in the antiferromagnetically ordered phase, forming two distinct
superconducting domes on the $P\!-\!T$ phase diagram, located around $P_{\rm max}$ and $P_{\rm
c}$.~\cite{kotegawa1,kotegawa2} Nuclear quadrupole resonance (NQR) measurements revealed that the
onset of the superconductivity is in both domes a consequence of the presence of the $4f$ electrons
of cerium.~\cite{harada1,harada2,harada3}

YNiGe$_3$ was used in our previous studies as an isostructural non-magnetic counterpart to
CeNiGe$_3$.~\cite{pikul} The temperature dependencies of its magnetic susceptibility, electrical
resistivity and specific heat (measured down to 1.7, 1.5 and 1.8~K, respectively) showed the
yttrium phase to be a simple metal with nearly temperature-independent magnetic susceptibility. In
the course of our investigations of the Ce$_{1-x}$Y$_x$NiGe$_3$ alloys (to be published elsewhere)
we have reinvestigated the electrical resistivity and the specific heat of YNiGe$_3$ down to
0.36~K. Here we report on a superconducting phase transition found in this compound.

\section{Material and methods}

Polycrystalline sample of YNiGe$_3$ was prepared by conventional arc melting the stoichiometric
amounts of the elemental components (Y 3N, Ni 3N, Ge 5N) in protective atmosphere of an argon glove
box. The pellet was subsequently wrapped in a molybdenum foil, sealed in an evacuated silica tube,
and annealed at 800$^{\circ}$C for one week. Quality of the product was verified by means of x-ray
powder diffraction and microprobe analysis, which both showed that the sample was a single phase.
Rietveld refinement of the x-ray powder pattern confirmed that YNiGe$_3$ crystallizes in the
SmNiGe$_3$ type structure with the lattice parameters $a$~=~4.060(1)~\AA, $b$~=~21.529(2)~\AA,
$c$~=~4.063(3)~\AA, being close to those reported previously.\cite{pikul} The energy dispersive
x-ray analysis of randomly chosen areas on the sample surface resulted in the average composition:
Y -- 25(1)at.\%, Ni -- 18(1)at.\% and Ge -- 57(1)at.\%, that corresponds to the formula
Y$_{1.3(1)}$Ni$_{0.9(1)}$Ge$_{2.8(1)}$. The deviation from the nominal composition may be
attributed to the fact, that the analysis was performed using internal standards of the
spectrometer.

Physical properties of the compound were studied using a commercial Quantum Design PPMS platform,
in temperature range 0.36--300~K and in magnetic fields up to 500~Oe, generated by a standard 9~T
magnet. The resistivity was measured by a conventional four point AC technique on a bar-shaped
sample with electrical contacts made of silver epoxy paste.

\section{Results and discussion}

\begin{figure}
\begin{center}
\includegraphics[width=\columnwidth]{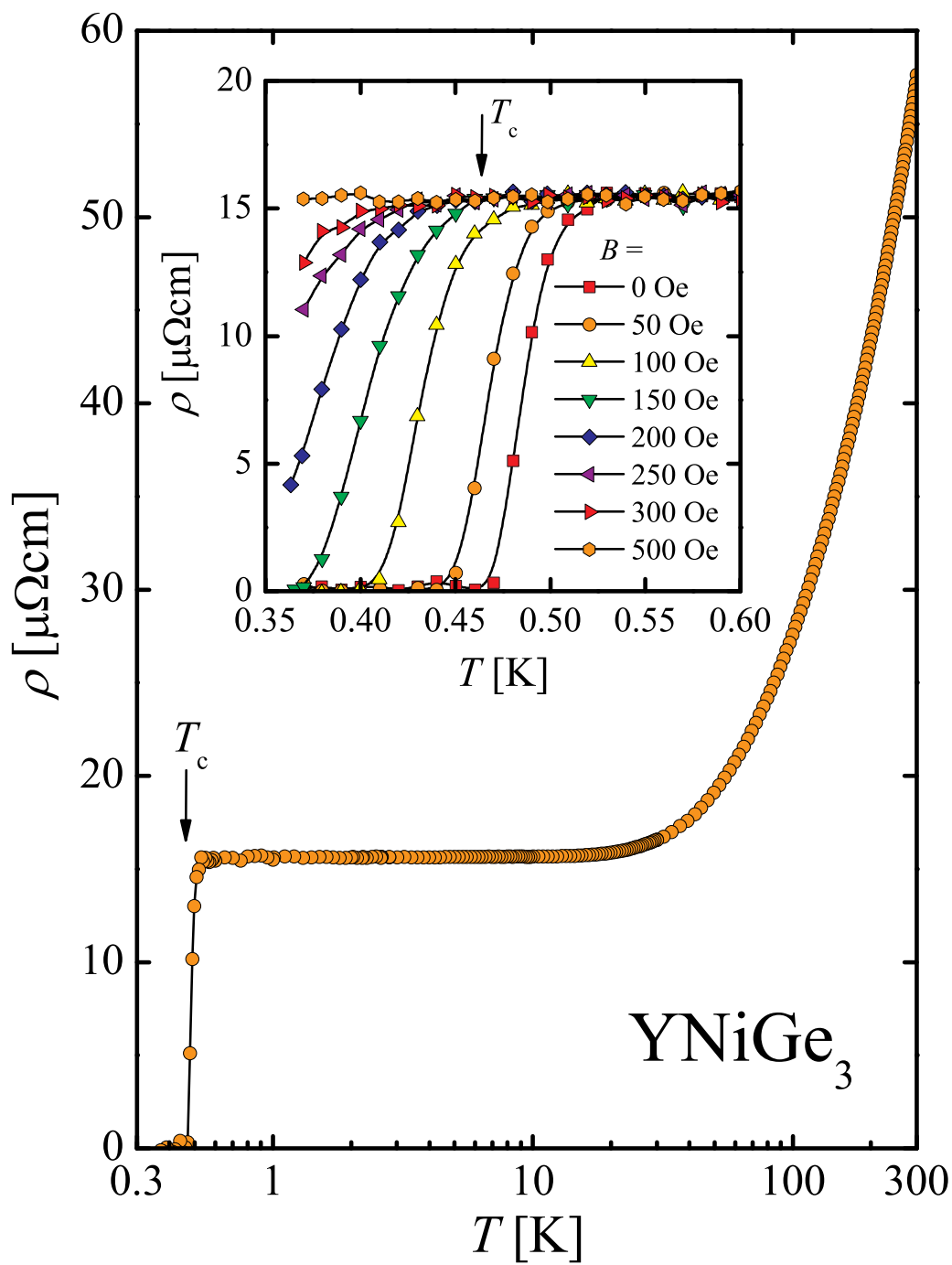}
\end{center}
\caption{Electrical resistivity of YNiGe$_3$ as a function of temperature.
The arrow marks the critical temperature $T_{\rm c}$ and the solid lines
serve as guides for the eye. The inset: evolution of the superconducting phase transition upon
applying magnetic field $B$; the values of $B$ are nominal.}
\label{f1}
\end{figure}

\begin{figure}
\begin{center}
\includegraphics[width=\columnwidth]{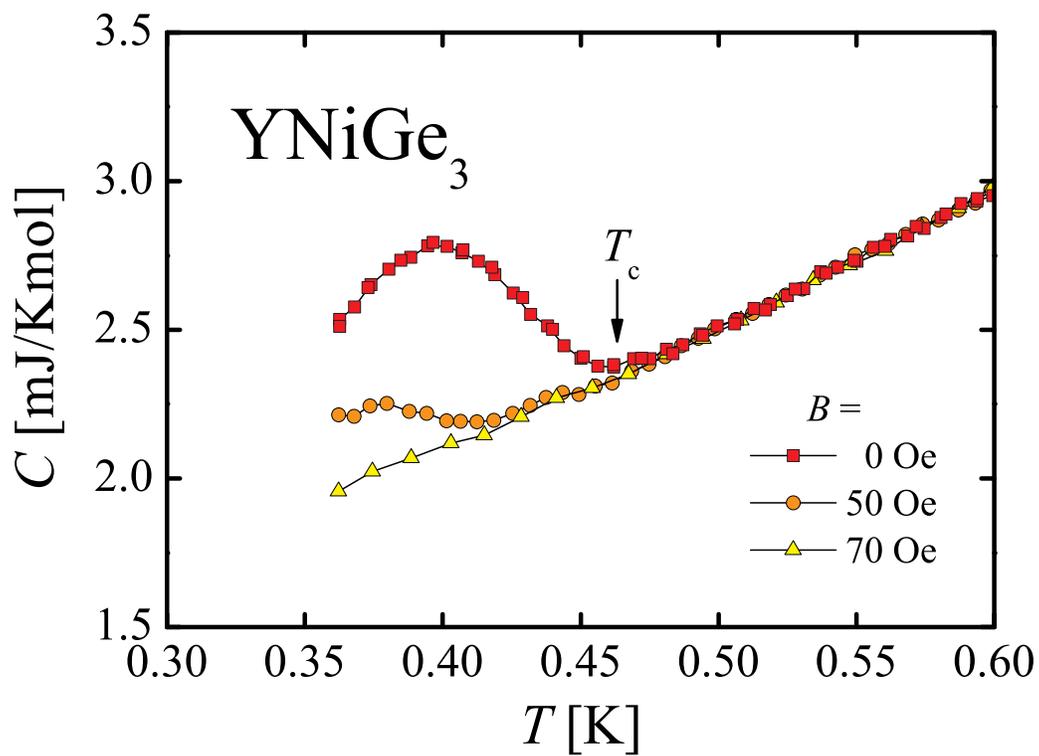}
\end{center}
\caption{Temperature dependence of the specific heat of YNiGe$_3$ measured in several different magnetic fields $B$;
the values of $B$ are nominal. The solid curves serve as guides for the eye and the arrow marks the
critical temperature.}
\label{f2}
\end{figure}

Figure~\ref{f1} presents the temperature dependence of the electrical resistivity $\rho$ of
YNiGe$_3$ in a semi-logarithmic scale. The overall behaviour of $\rho(T)$ is similar to that
reported previously:~\cite{pikul} upon cooling down, $\rho$ decreases in a simple metallic manner
and saturates below about 10~K. The resistivity remains constant down to the temperature of about
0.50~K, below which a rapid drop of $\rho$ clearly manifests the onset of superconductivity, with
the critical temperature $T_{\rm c}$~=~0.46~K. Small width of the phase transition, which is less
than 8\% of $T_{\rm c}$, points at a bulk character of the superconductivity. Presence of a
distinct anomaly, that appears in the temperature variation of the specific heat $C$ of YNiGe$_3$
just below $T_{\rm c}$~=~0.46~K (Fig.~\ref{f2}), seems to support the presumption on the intrinsic
nature of the superconducting state. Nevertheless, low-temperature magnetic susceptibility
measurements are needed to unambiguously confirm the bulk character of the superconducting state in
the compound studied.

As seen in the insets to Figs.~\ref{f1} and \ref{f2}, the superconducting state in YNiGe$_3$ is
very sensitive on the strength of the magnetic field applied. Upon increasing field, $T_{\rm c}$
inferred from the $\rho(T)$ curve systematically decreases and already at 500~Oe the drop of $\rho$
is out of the investigated temperature range. Similar behaviour is visible in the specific heat,
yet the anomaly in $C(T)$ quickly broadens with increasing $B$ and therefore it is hardly visible
on the experimental curve already in 70~Oe. Due to low resolution of the magnet used in our
experiments (the remanent field was of the order of 10--20 Oe), we were not able to construct and
reliably analyse the $B\!-\!T$ phase diagram for YNiGe$_3$. Nevertheless, based on the electrical
resistivity measurements one can estimate the critical field value $H_{c2}$ to be of the order of
500~Oe.

The small value of $H_{\rm c2}$ is characteristic of conventional, BCS-like superconductors.
However, the observed jump of the specific heat, $\Delta C / (\gamma_n T_{\rm c})$, is in YNiGe$_3$
of about 0.50(5) (Fig.~\ref{f2}), which is well below $\Delta C / (\gamma_n T_{\rm c})$~=~1.43
expected for an $s$-wave BCS superconductor. Such a distinct deviation of the specific heat jump
from the BCS value hints at possible unconventional character of the superconducting state. For
example, in YBa$_2$Cu$_3$O$_7$, which is one of the best known high-$T_{\rm c}$ superconductors,
$\Delta C / (\gamma_n T_{\rm c})$ is of about 2,\cite{plackowski} while in the multi-valued gap
superconductor MgB$_2$ the jump in $C(T)$ is only 0.82.\cite{plackowski} The recently reported
noncentrosymmetric strongly-coupled superconductor Mo$_3$Al$_2$C has $\Delta C / (\gamma_n T_{\rm
c})$ equal to 2.28.\cite{bauer}

\section{Conclusions}

The compound YNiGe$_3$ was found to exhibit an ambient-pressure superconductivity below $T_{\rm
c}$~=~0.46~K, which is close to the maximal $T_{\rm c}$ evidenced in CeNiGe$_3$ around the quantum
critical point.\cite{nakashima1,nakashima2,kotegawa1,kotegawa2} In this context it is worth noting
that substitution of Ce by Y results in 6\%-reduction of the unit cell volume of the system, which
can be roughly considered as equivalent to high hydrostatic pressure. Therefore, the finding of the
superconductivity in YNiGe$_3$, which does not contain $f$-electrons, may shed new light on the
character of the superconductivity in the isostructural CeNiGe$_3$ compound.

\section*{Acknowledgements}

This work was supported by the Polish Ministry of Sciences and Higher Education through the
research grant No.~N~N202~102338. The authors thank K.~Wochowski, R.~Gorzelniak and D.~Badurski for
their assistance in performing the experiments.

\end{document}